\documentclass[lettersize,journal]{IEEEtran}

\usepackage{amsmath,amsfonts,amssymb}
\usepackage{algorithmic}
\usepackage[ruled,vlined]{algorithm2e}
\usepackage{array}
\usepackage{tabularx}
\usepackage[caption=false,font=footnotesize,labelfont=rm,textfont=rm]{subfig}
\usepackage{textcomp}
\usepackage{stfloats}
\usepackage{url}
\usepackage{verbatim}
\usepackage{graphicx}
\usepackage{float}
\usepackage[table]{xcolor}
\usepackage{multirow}
\usepackage{makecell}
\usepackage{pifont}
\usepackage{booktabs}
\usepackage{arydshln}
\usepackage{bigstrut}
\usepackage{microtype}
\usepackage{tikz}
\usepackage{mdframed}
\usetikzlibrary{arrows.meta,positioning,calc,backgrounds}
\usepackage[hidelinks]{hyperref}

\graphicspath{{./}}
\def\BibTeX{{\rm B\kern-.05em{\sc i\kern-.025em b}\kern-.08em
    T\kern-.1667em\lower.7ex\hbox{E}\kern-.125emX}}
\usepackage{balance}

\newcommand{\meanstd}[2]{#1$_{\pm #2}$}

\newmdenv[
  backgroundcolor=black!3,
  linecolor=black,
  linewidth=0.65pt,
  innerleftmargin=8pt,
  innerrightmargin=8pt,
  innertopmargin=7pt,
  innerbottommargin=7pt,
  skipabove=0pt,
  skipbelow=0pt
]{promptframe}



\RestyleAlgo{ruled}

\begin{document}

\title{LLaTSA: Large Language Model-Aligned General-Purpose Transient Stability Analysis}

\author{Chao Shen,~\IEEEmembership{Student Member,~IEEE},
        Hongwei Zhen,~\IEEEmembership{Student Member,~IEEE}, Junyan Shao,~\IEEEmembership{Member,~IEEE},
        Zhenghao Yang,~\IEEEmembership{Student Member,~IEEE},
        Yifan Zhang,~\IEEEmembership{Student Member,~IEEE},
        Mingyang Sun,~\IEEEmembership{Senior Member,~IEEE}%
}

\markboth{XXX}%
{LLaTSA: Large Language Model-Aligned General-Purpose Transient Stability Analysis}

\maketitle

\begin{abstract}
    Dynamic trajectory prediction has become an important paradigm for data-driven transient stability analysis (TSA), yet most existing predictors remain system-specific and require substantial retraining when network configurations, generation mixes, or state-variable sets change. Uni-TSA introduced a general-purpose TSA framework that combines channel-independent modeling with a pretrained large language model (LLM) predictor. Nevertheless, its application to heterogeneous systems is limited by ambiguity in short observations, a mismatch between numerical trajectories and LLM embeddings, neglected coupling among state variables, and the high inference cost of dense backbones. This paper proposes LLaTSA, an LLM-aligned framework for general-purpose trajectory-based TSA. LLaTSA first incorporates operating conditions, disturbance attributes, and state-variable identity through a structured textual prefix. It then aligns normalized temporal patches with a TSA-related vocabulary before processing them with a pretrained sparse decoder-only mixture-of-experts (MoE) backbone. A state-variable coupling module captures coordinated post-fault evolution, while teacher forcing and rollout-based training support iterative long-horizon prediction. Case studies on multiple test systems demonstrate accurate trajectory prediction, reliable stability discrimination, and effective adaptation across unseen scenarios.
    \end{abstract}


\section{Introduction}
{\color{black}
The growing penetration of renewable generation and converter-connected devices has increased the uncertainty and dynamic complexity of modern power systems \cite{hatziargyriou2020definition}. Compared with conventional synchronous generators, converter-connected resources introduce different dynamic response characteristics, which can affect system inertia, damping, and post-disturbance stability behavior \cite{shen2025physics}. Transient stability analysis (TSA) evaluates post-disturbance rotor angle dynamics to determine whether the system can maintain synchronism following severe disturbances \cite{shen2025universal, zuo2025probabilistic}. Accurate and computationally efficient TSA is therefore essential for secure system operation in online security analysis \cite{huang2019deep}, preventive control \cite{li2026spatially}, and emergency decision support \cite{liu2022real}.

Conventional TSA methods can be broadly divided into time-domain simulation (TDS) and direct methods. TDS evaluates transient stability by numerically solving the system differential-algebraic equations under specified disturbances and observing the resulting post-disturbance trajectories. Although this procedure provides accurate dynamic responses, it is computationally expensive for large-scale screening over multiple OCs and contingency scenarios \cite{xia2018efficient}. Direct methods seek to assess stability without explicit trajectory simulation, typically through analytical stability criteria \cite{xue1989extended}, transient energy functions \cite{chang1995direct}, or Lyapunov-based formulations \cite{qiu2022adaptive}. However, their scalability and robustness in practical systems are often limited by model dependence, conservatism, and implementation complexity. These limitations have led to growing interest in data-driven TSA for fast security assessment.

Data-driven TSA has mainly developed along two directions: stability assessment and dynamic trajectory prediction. Stability assessment methods use post-fault measurements to infer stability labels, transient stability margins, or security indices. Early classifiers based on artificial neural networks \cite{zhou1994application} and support vector machines \cite{gomez2010support} were later extended by attention-based models \cite{chen2022interpretable}, convolutional architectures \cite{shi2020convolutional}, margin estimation methods \cite{su2023online}, and security-index prediction models \cite{zhu2019hierarchical}. These methods provide fast stability decisions but compress the post-disturbance response into labels or scalar indices. 
Dynamic trajectory prediction preserves the temporal structure of post-fault dynamics by predicting future state trajectories from operating conditions, disturbance information, and early trajectory observations. The predicted trajectories typically include rotor angles, rotor speeds, bus voltages, or other state channels relevant to TSA. This formulation provides time-resolved dynamic information and can support both stability assessment and subsequent control analysis \cite{bahbah2004new}. Representative methods include physics-informed predictors \cite{misyris2020physics, shen2025physics}, recurrent sequence models \cite{ye2024use}, graph-based models \cite{zhao2022structure}, frequency-domain frameworks \cite{cui2023frequency}, and probabilistic predictors \cite{tan2025bayesian}. Despite these advances, most existing trajectory-prediction methods assume fixed system configurations. As power systems evolve through network expansion, generation portfolio changes, and converter integration, existing models often require data-intensive redesign and computationally expensive retraining to accommodate changes in system size, topology, and dynamic component composition.

To improve cross-system applicability, Uni-TSA \cite{shen2025universal} introduced general-purpose trajectory-based TSA. Its channel-independent formulation mitigates input-output dimensional mismatch across systems with different numbers of state variables, and its dense LLM backbone provides a shared temporal predictor. However, reliable online application across heterogeneous systems remains limited for four reasons. First, short post-fault observations may not distinguish cases with divergent future responses. Second, directly projecting numerical trajectories into a text-pretrained embedding space leaves a modality mismatch. Third, channel-wise prediction omits inter-variable dependence in post-fault dynamics. Finally, scaling dense LLM backbones markedly prolongs inference time and limits their suitability for online TSA. These limitations are analyzed in \autoref{sec:limitations}.}

{\color{black}To address these challenges, this paper proposes \textit{LLaTSA}, a LLM-aligned framework for general-purpose trajectory-based TSA. LLaTSA combines scenario-aware textual conditioning, vocabulary-guided trajectory alignment, sparse mixture-of-experts (MoE) temporal modeling, state-variable coupling, and rollout-based training. The main contributions are as follows:

\begin{enumerate}
    \item To the best of the authors' knowledge, LLaTSA is the first LLM-aligned framework for general-purpose trajectory-based TSA. It jointly addresses short-observation ambiguity, missing state-variable coupling, numerical--textual representation mismatch, and the capacity--efficiency tradeoff of dense backbones.

    \item A scenario-aware, vocabulary-guided trajectory encoder is developed. It combines structured power-system operating and disturbance context with cross-attention alignment between normalized temporal patches and TSA-related vocabulary embeddings, allowing the pretrained backbone to jointly model contingency information and post-fault dynamic responses.

    \item A pretrained sparse decoder-only MoE backbone is integrated with a state-variable coupling module for post-fault power-system trajectory prediction. It combines top-$K$ expert routing of aligned temporal tokens with cross-variable coupling of their representations, allowing efficient modeling of post-fault dynamics and their inter-variable dependencies.

    \item Teacher forcing and rollout-based training are combined for iterative prediction. Results across multiple test systems demonstrate accurate trajectory prediction, reliable stability discrimination, and the generalization and interpretability of LLaTSA under heterogeneous power-system dynamics.
\end{enumerate}}

\begin{figure}
  \centering
  \includegraphics[width=\linewidth]{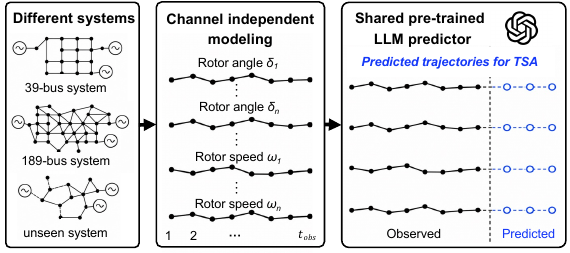}
  \caption{Uni-TSA framework based on channel-independent trajectory modeling.}
  \label{fig:unitsa_framework}
\end{figure}

{\color{black}
\section{Preliminaries and General-Purpose TSA Formulation}

\subsection{Power System Transients and Data-driven TSA Paradigms}

The dynamics of a power system are described by nonlinear differential-algebraic equations (DAEs) that couple device dynamics and network algebraic constraints \cite{shen2025physics}:
\begin{align}
\dot{\mathbf{x}}(t) &= \mathbf{f}\big(\mathbf{x}(t), \mathbf{y}(t), \mathbf{d}(t), t\big), \\
\mathbf{0} &= \mathbf{g}\big(\mathbf{x}(t), \mathbf{y}(t), \mathbf{d}(t), t\big),
\label{eq:dae_model}
\end{align}
where $\mathbf{x}(t)\in\mathbb{R}^{n_x}$ contains dynamic states such as generator rotor angles $\boldsymbol{\delta}(t)$, $\mathbf{y}(t)\in\mathbb{R}^{n_y}$ contains algebraic variables such as bus voltage magnitudes $\mathbf{V}(t)$, and $\mathbf{d}(t)\in\mathbb{R}^{n_d}$ denotes the disturbance or control input. TDS obtains post-disturbance trajectories by numerically integrating \eqref{eq:dae_model} for each OC and contingency. The resulting trajectories can be used to determine stability through criteria such as the transient stability index (TSI):
\begin{equation}
S=
\begin{cases}
\mathrm{stable}, & \mathrm{TSI}\ge 0,\\
\mathrm{unstable}, & \mathrm{TSI}<0,
\end{cases}
\qquad
\mathrm{TSI}=\frac{\delta_T-|\Delta\delta|_{\max}}{\delta_T+|\Delta\delta|_{\max}},
\end{equation}
where $|\Delta\delta|_{\max}$ is the maximum rotor angle separation and $\delta_T$ is the stability threshold. The repeated numerical integration required by TDS is costly for large-scale online screening. Data-driven stability classification improves efficiency by mapping system data directly to stability labels or security indices \cite{su2023online, zhu2019hierarchical, shen2025physics_augmented}, but it compresses the post-disturbance response into scalar outputs.

Dynamic-trajectory-prediction-based TSA methods learn a data-driven mapping from observed post-disturbance trajectory segments to future state evolution. Let
\begin{equation}
\mathbf{X}_{1:T}=[\mathbf{x}_1,\mathbf{x}_2,\ldots,\mathbf{x}_T]^\top \in \mathbb{R}^{T\times n_x},
\label{eq:state_prefix}
\end{equation}
denote an observed post-disturbance trajectory segment. Given a prediction model $\Phi(\cdot;\Theta)$ and horizon $H$, the future trajectory is expressed as
\begin{equation}
\widehat{\mathbf{X}}_{T+1:T+H}
=
\Phi\big(\mathbf{X}_{1:T};\Theta\big),
\label{eq:system_multistep_prediction}
\end{equation}
where $\widehat{\mathbf{X}}_{T+1:T+H}$ can be produced by direct multi-step prediction \cite{shen2025physics,ye2024use} or iterative rollout \cite{zhao2022structure}. Compared with stability classification, trajectory prediction preserves time-resolved dynamic information for subsequent stability and control analysis.


\subsection{General-Purpose Trajectory Prediction for TSA}

As power systems expand and integrate increasing shares of converter-connected resources, TSA must cover a wider range of network configurations, disturbance scenarios, and OCs with limited retraining. This requirement defines the general-purpose TSA setting considered in this paper. Existing trajectory-prediction methods face three obstacles to achieving this generality. First, \textit{data-processing rigidity} arises because system-specific mappings $\Phi(\mathbf{X}_{1:T};\Theta)$ tie $\mathbf{X}_{1:T}\in\mathbb{R}^{T\times n_x}$ to a fixed set and ordering of state variables. Changes in the number of SGs, measured buses, or state channels alter $n_x$ and often require model redefinition. Second, \textit{architectural constraints} arise because compact recurrent predictors such as LSTM and GRU are usually tailored to a specific system and disturbance set. Their reuse is therefore limited for heterogeneous rotor-angle and rotor-speed responses. Third, \textit{error propagation} can accumulate during short-window online rollout and distort rotor-angle separation or rotor-speed recovery over the prediction horizon.

Uni-TSA \cite{shen2025universal} addresses these challenges through channel-independent trajectory modeling, as shown in \autoref{fig:unitsa_framework}. The system-level trajectory $\mathbf{X}_{1:T}\in\mathbb{R}^{T\times n_x}$ is decomposed into univariate state-channel sequences,
$\mathbf{X}_{1:T}\rightarrow \{\mathbf{x}_{1:T}^{(j)}\}_{j=1}^{n_x}$,
which are processed by a shared predictor independent of the number and ordering of state variables. A pretrained LLM backbone is used for temporal prediction, and teacher-forcing/scheduled-sampling fine-tuning is adopted to reduce rollout error accumulation. Reported results show zero-shot generalization to unseen faults and mixed-stability conditions and few-shot adaptation to a heterogeneous system with 5\% target-system data.}

\begin{figure}
    \centering
    \subfloat[Short-observation ambiguity.\label{fig:short_observation_ambiguity}]{
        \includegraphics[width=0.46\linewidth]{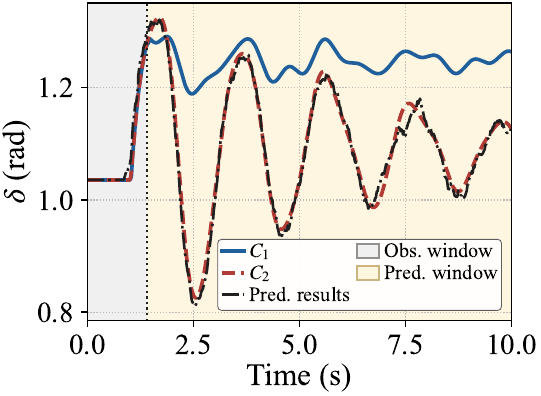}
    }
    \subfloat[Inter-variable correlation.\label{fig:inter_variable_correlation}]{
        \includegraphics[width=0.46\linewidth]{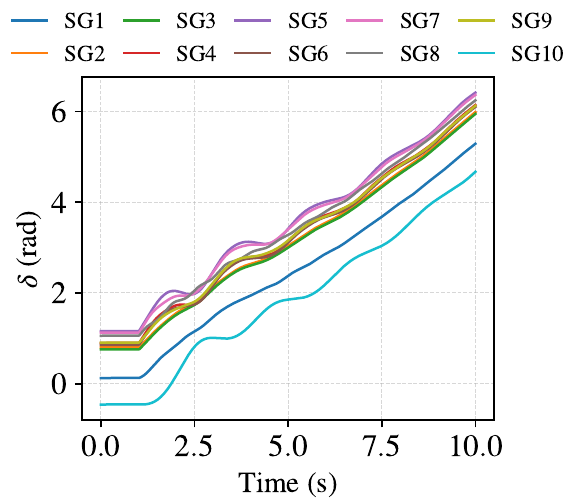}
    }\\[-1mm]
    \subfloat[Modality gap.\label{fig:modality_gap}]{
        \includegraphics[width=0.46\linewidth]{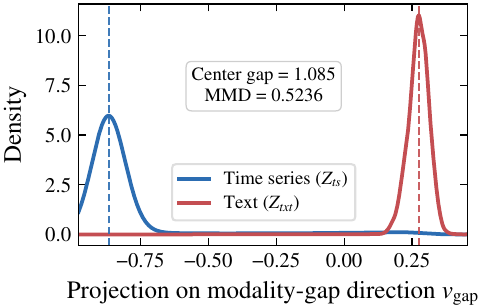}
    }
    \subfloat[Capacity-efficiency tradeoff.\label{fig:unitsa_capacity_efficiency_tradeoff}]{
        \includegraphics[width=0.46\linewidth]{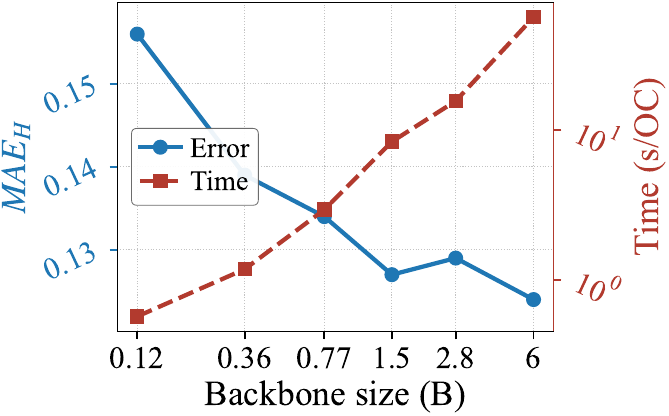}
    }
    \caption{Evidence for Uni-TSA limitations. (a) Similar short-window observations can lead to different future trajectories. (b) Post-fault rotor angle responses exhibit strong inter-variable correlation. (c) Dynamic trajectory embeddings and textual embeddings show clear distributional separation along the modality gap direction in the shared representation space. (d) Dense-backbone scaling reduces prediction error with rapidly increased inference time.}
    \label{fig:limitation_evidence}
\end{figure}

{\color{black}
\section{LLaTSA Framework: Design and Formulation}

\subsection{Limitations of Uni-TSA for General-Purpose TSA}
\label{sec:limitations}

Uni-TSA provides an important baseline for general-purpose trajectory-based TSA, but its application across heterogeneous power systems remains constrained by several unresolved limitations, including short-observation discrimination, inter-variable coupling, cross-modal representation mismatch, and dense-backbone computation. This subsection analyzes these limitations and motivates the design of LLaTSA.

\subsubsection{Insufficient Discrimination Under Short Observations}\label{sec:short_observation_discrimination}

Uni-TSA decomposes multivariable post-fault trajectories into univariate state-channel sequences and predicts each channel from its own observation history. This channel-wise formulation may be ambiguous under short post-fault observation windows, where different operating-disturbance cases can exhibit similar early responses. As illustrated in \autoref{fig:limitation_evidence}(a), let $C_1$ and $C_2$ denote two operating-disturbance cases. For the $j$-th state channel, their early observations may satisfy
$\| \mathbf{x}_{1:T}^{(j)}(C_1) - \mathbf{x}_{1:T}^{(j)}(C_2) \|_2 \le \eta$,
where $\eta$ is a small positive constant, while their future trajectories diverge because of different operating conditions or disturbance attributes. Therefore, prediction based only on the short univariate prefix may fail to distinguish cases with different subsequent dynamic responses, thereby degrading long-horizon trajectory accuracy.

\subsubsection{Weak Representation of Coupled State Evolution}\label{sec:coupled_state_evolution}

Post-fault trajectories in TSA usually contain correlated variations among state variables, whereas Uni-TSA predicts each channel separately $\mathbf{X}_{1:T}\rightarrow \{\mathbf{x}_{1:T}^{(j)}\}_{j=1}^{n_x}$. This univariate treatment reduces the dependence on a fixed state dimension, but omits an explicit representation of correlations among state variables. As shown in \autoref{fig:limitation_evidence}(b), the rotor angle trajectories of different synchronous generators (SGs) exhibit highly consistent post-fault trends in a stable case, with pairwise correlation coefficients above $0.9$. This correlation indicates that post-fault state variables evolve in a coupled manner. Treating each variable as an independent sequence may lead to multivariable predictions that are inconsistent with the coordinated generator dynamics.

\subsubsection{Modality Gap Between Numerical Trajectories and Textual Representations}\label{sec:numerical_trajectory_mismatch}

Uni-TSA employs a pretrained LLM as the sequence-modeling backbone for post-fault trajectory prediction. The backbone is pretrained on textual sequences, whereas post-fault trajectories are continuous numerical records of dynamic responses, including rotor angle, rotor speed, and voltage variations. This difference raises a representation-consistency issue: trajectory information relevant to TSA may be poorly organized by a backbone whose pretrained structure is derived from text.
To examine this issue, we compare the dynamic trajectory embeddings $\mathbf{Z}_{\mathrm{ts}}$ with the textual embeddings $\mathbf{Z}_{\mathrm{txt}}$ in the original high-dimensional embedding space. After $L_2$ normalization, both modalities are projected onto the direction between their modality centers, i.e., $\mathbf{v}_{\mathrm{gap}}=(\boldsymbol{\mu}_{\mathrm{txt}}-\boldsymbol{\mu}_{\mathrm{ts}})/\|\boldsymbol{\mu}_{\mathrm{txt}}-\boldsymbol{\mu}_{\mathrm{ts}}\|_2$. As shown in \autoref{fig:limitation_evidence}(c), the projected density distributions are clearly separated, with a center gap of 1.085 and a maximum mean discrepancy (MMD) of 0.5236. This result indicates a systematic modality gap in the shared embedding space rather than a visual artifact caused by low-dimensional plotting, which can limit the direct reuse of a backbone pretrained on text for post-fault trajectory modeling.

\subsubsection{Capacity-Efficiency Tradeoff in Dense Backbones}\label{sec:dense_backbone_capacity_efficiency}

Uni-TSA adopts a dense pretrained LLM backbone to enhance the representation capacity of channel-independent trajectory prediction. However, dense scaling becomes computationally restrictive when a larger backbone is used to cover more diverse post-fault dynamic responses across operating conditions, disturbance scenarios, and system configurations. In a dense Transformer, all feed-forward network (FFN) parameters are activated for every trajectory token. Therefore, increasing the number of Transformer layers $L$ or the hidden dimension $d_{\mathrm{model}}$ increases the activated FFN computation per token, which scales approximately as $\mathcal{O}(L d_{\mathrm{model}}^2)$ when the FFN width grows with $d_{\mathrm{model}}$.
As shown in \autoref{fig:limitation_evidence}(d), replacing the Uni-TSA backbone with larger dense LLMs reduces trajectory prediction error, but the error reduction is not proportional to the increase in model size. Meanwhile, complete-trajectory inference time increases rapidly because all FFN parameters are evaluated for each token in the larger dense backbone. This capacity-efficiency tradeoff constrains direct scaling toward larger dense models, which are desirable for representing broader classes of post-fault dynamics but may be unsuitable for online TSA with limited time for stability assessment and emergency-control screening.}

To address the above limitations, this paper proposes LLaTSA with four coordinated designs: scenario-aware textual conditioning, vocabulary-guided trajectory alignment, sparse MoE backbone modeling, and state variable coupling. Textual conditioning incorporates OC and disturbance information for short-observation prediction. Trajectory alignment mitigates the modality gap between numerical trajectories and the representation space learned from text pretraining. Sparse MoE modeling mitigates the capacity-efficiency tradeoff of dense scaling by activating only a subset of experts per token. State variable coupling represents the dependence among post-disturbance state variables for multivariable trajectory prediction. The corresponding formulations are given in the following subsections.

{\color{black}
\subsection{Dynamic Trajectory Encoding and Representation Alignment}
\label{sec:trajectory_alignment}

\begin{figure}
    \centering
    \includegraphics[width=\linewidth]{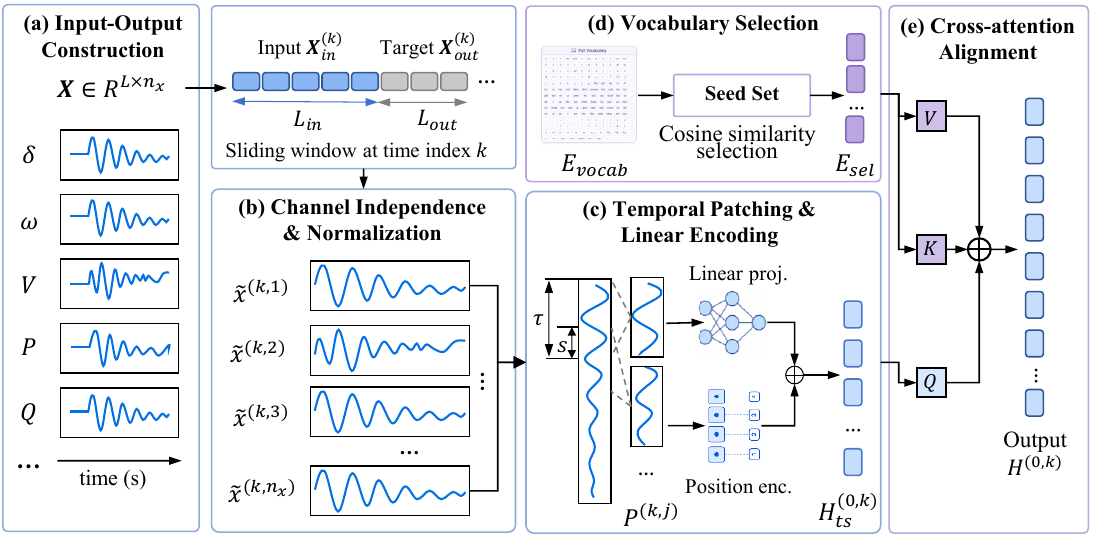}
    \caption{State-trajectory encoding and vocabulary-guided alignment. Post-fault multivariable trajectories are split into input and target windows, decomposed into normalized univariate sequences, encoded as temporal patches, and aligned with selected TSA-related vocabulary embeddings through cross-attention.}
    \label{fig:encoding_alignment}
\end{figure}

This subsection presents the trajectory-encoding procedure that maps post-disturbance dynamic sequences to the pretrained LLM backbone. Since the LLM backbone processes fixed-dimensional input vectors, numerical dynamic trajectories are first mapped to the required model dimension. Uni-TSA uses linear projection for this mapping. However, the analysis in \autoref{sec:numerical_trajectory_mismatch} shows that dynamic trajectory embeddings and textual embeddings remain separated in the shared representation space, which weakens the direct reuse of the backbone pretrained on text for numerical trajectory prediction.

LLaTSA combines sequence encoding with vocabulary-guided alignment. Post-disturbance trajectories are decomposed into univariate state-variable sequences, normalized, and segmented into temporal patches to form ordered trajectory representations. These representations are aligned with selected pretrained token representations, so that numerical dynamic information is moved toward a textual representation basis relevant to TSA. The four-step procedure is shown in \autoref{fig:encoding_alignment} and detailed below.

\subsubsection{Dynamic Response Input-Output Pair Construction}

Let $\mathbf{X}\in\mathbb{R}^{L\times n_x}$ denote a post-disturbance state trajectory of length $L$. Input-output pairs are extracted from $\mathbf{X}$ using a sliding-window strategy as shown in \autoref{fig:encoding_alignment}(a). Specifically, given an input window length $L_{\mathrm{in}}$ and an output window length $L_{\mathrm{out}}$, the $k$-th sample is defined as
\begin{equation}
\mathbf{X}^{(k)}_{\mathrm{in}}
=
[\mathbf{x}_k,\mathbf{x}_{k+1},\ldots,\mathbf{x}_{k+L_{\mathrm{in}}-1}]^T,
\end{equation}
\begin{equation}
\begin{aligned}
\mathbf{X}^{(k)}_{\mathrm{out}}
=
\big[
&\mathbf{x}_{k+L_{\mathrm{in}}},
\mathbf{x}_{k+L_{\mathrm{in}}+1},\ldots,\mathbf{x}_{k+L_{\mathrm{in}}+L_{\mathrm{out}}-1}
\big]^T .
\end{aligned}
\end{equation}
Here, $\mathbf{X}^{(k)}_{\mathrm{in}}\in\mathbb{R}^{L_{\mathrm{in}}\times n_x}$ corresponds to the available post-fault observation window, and $\mathbf{X}^{(k)}_{\mathrm{out}}\in\mathbb{R}^{L_{\mathrm{out}}\times n_x}$ denotes the subsequent dynamic trajectory to be predicted.

\subsubsection{State-Variable Decomposition and Normalization}
The input sample $\mathbf{X}^{(k)}_{\mathrm{in}}$ is first decomposed along the variable dimension, as shown in \autoref{fig:encoding_alignment}(b), so that the shared encoder can process systems with different numbers of state variables. The trajectory of the $j$-th state variable is extracted as
\begin{equation}
\mathbf{x}^{(k,j)}
=
[x^{(j)}_k,x^{(j)}_{k+1},\ldots,x^{(j)}_{k+L_{\mathrm{in}}-1}]^\top
\in\mathbb{R}^{L_{\mathrm{in}}}.
\end{equation}
Each state-variable trajectory is then independently normalized to reduce scale differences among heterogeneous variables:
\begin{equation}
\widetilde{\mathbf{x}}^{(k,j)}
=
\frac{\mathbf{x}^{(k,j)}-\mu^{(k,j)}}{\sigma^{(k,j)}+\epsilon},
\end{equation}
where $\mu^{(k,j)}$ and $\sigma^{(k,j)}$ are the mean and standard deviation computed from the input trajectory segment $\mathbf{x}^{(k,j)}$ itself, and $\epsilon$ is a small constant for numerical stability. The normalized trajectories $\{\widetilde{\mathbf{x}}^{(k,j)}\}_{j=1}^{n_x}$ are then processed by the shared encoder.

\subsubsection{Temporal Segmentation and Linear Encoding}
The normalized trajectory $\widetilde{\mathbf{x}}^{(k,j)}$ is divided into overlapping temporal segments with length $\tau$ and stride $s$, as illustrated in \autoref{fig:encoding_alignment}(c). The $i$-th segment of the $j$-th state variable in the $k$-th sample is defined as
\begin{equation}
\begin{aligned}
\mathbf{p}^{(k,j)}_i
=
\big[
&\widetilde{x}^{(k,j)}_{(i-1)s+1},
\widetilde{x}^{(k,j)}_{(i-1)s+2},\ldots,\\
&\widetilde{x}^{(k,j)}_{(i-1)s+\tau}
\big]^\top ,
\end{aligned}
\end{equation}
where $\mathbf{p}^{(k,j)}_i\in\mathbb{R}^{\tau}$ and $i=1,\ldots,N_p$. The number of temporal segments is
\begin{equation}
N_p=\left\lfloor\frac{L_{\mathrm{in}}-\tau}{s}\right\rfloor+1 .
\end{equation}
By stacking these segments, the sequence representation of the $j$-th state variable is written as
\begin{equation}
\mathbf{P}^{(k,j)}
=
[\mathbf{p}^{(k,j)}_1;\mathbf{p}^{(k,j)}_2;\ldots;\mathbf{p}^{(k,j)}_{N_p}]
\in\mathbb{R}^{N_p\times\tau}.
\end{equation}
The segment sequences of all state variables are stacked as
\begin{equation}
\mathbf{P}^{(k)}
=
\big[
\mathbf{P}^{(k,1)},\mathbf{P}^{(k,2)},\ldots,\mathbf{P}^{(k,n_x)}
\big]^\top
\in\mathbb{R}^{n_x\times N_p\times\tau}.
\end{equation}

Each temporal segment is then mapped to the model dimension through a learnable linear transformation:
\begin{equation}
\mathbf{E}^{(k)}_{\mathrm{ts}}
=
\mathbf{P}^{(k)}\cdot\mathbf{W}_{\mathrm{ts}}+\mathbf{b}_{\mathrm{ts}}
\in\mathbb{R}^{n_x\times N_p\times d_{\mathrm{model}}},
\end{equation}
where $\mathbf{W}_{\mathrm{ts}}\in\mathbb{R}^{\tau\times d_{\mathrm{model}}}$ and $\mathbf{b}_{\mathrm{ts}}\in\mathbb{R}^{d_{\mathrm{model}}}$ are trainable parameters, and $d_{\mathrm{model}}$ is the hidden dimension of the pretrained LLM backbone. Positional encoding is added to preserve the temporal order:
\begin{equation}
\mathbf{H}^{(0,k)}_{\mathrm{ts}}
=
\mathbf{E}^{(k)}_{\mathrm{ts}}+\mathbf{P}_{\mathrm{pos}}
\in\mathbb{R}^{n_x\times N_p\times d_{\mathrm{model}}},
\end{equation}
where $\mathbf{P}_{\mathrm{pos}}\in\mathbb{R}^{N_p\times d_{\mathrm{model}}}$ denotes the positional encoding matrix.
The patch count $N_p$ may vary with the system-specific observation length $L_{\mathrm{in}}$. The corresponding positional embeddings are selected within the backbone context limit. The resulting $\mathbf{H}^{(0,k)}_{\mathrm{ts}}$ contains ordered representations of local post-disturbance trajectory variations for all state variables.

\subsubsection{State-Trajectory Vocabulary Alignment}

LLaTSA then aligns $\mathbf{H}^{(0,k)}_{\mathrm{ts}}$ with selected entries in the pretrained LLM vocabulary. This design follows from the input structure of the LLM backbone, whose native inputs are defined through the vocabulary embedding table, whereas $\mathbf{H}^{(0,k)}_{\mathrm{ts}}$ is derived from dynamic trajectories and may remain separated from textual representations in the shared embedding space. Directly considering the full vocabulary is inefficient and may introduce tokens unrelated to TSA. Therefore, LLaTSA constructs a compact vocabulary subset from TSA-related terms and defines the alignment basis through the selected vocabulary entries, as shown in \autoref{fig:encoding_alignment}(d)(e). The alignment procedure is described below.

\noindent\textit{(a) Literature-guided vocabulary selection.}
A TSA-related seed term set $\mathcal{S}=\{s_r\}_{r=1}^{R}$ is collected from trajectory-prediction and transient-stability literature \cite{misyris2020physics, shen2025physics, ye2024use, zhao2022structure, tan2025bayesian, shen2025universal}, covering representative concepts such as stability status, fault type, clearing time, rotor angle deviation, oscillation, and recovery. Tokenizing these seed terms with the pretrained backbone tokenizer gives
\begin{equation}
\mathcal{U}_{\mathcal{S}}
=
\bigcup_{r=1}^{R}\mathcal{T}(s_r),
\end{equation}
where $\mathcal{T}(s_r)$ denotes the token indices of seed term $s_r$. The compact vocabulary subset is then obtained by retrieving, from the frozen LLM vocabulary embedding table, the top-$m$ tokens nearest to each seed-token embedding:
\begin{equation}
\mathcal{V}_{\mathrm{sel}}
=
\bigcup_{u\in\mathcal{U}_{\mathcal{S}}}
\mathrm{Top}\text{-}m_{v\in\mathcal{V}}
\cos\big(\mathbf{E}_{\mathrm{vocab}}[v,:],\mathbf{E}_{\mathrm{vocab}}[u,:]\big).
\end{equation}
The selected embedding table $\mathbf{E}_{\mathrm{sel}}\in\mathbb{R}^{|\mathcal{V}_{\mathrm{sel}}|\times d_{\mathrm{model}}}$ is obtained by indexing $\mathbf{E}_{\mathrm{vocab}}\in\mathbb{R}^{|\mathcal{V}|\times d_{\mathrm{model}}}$ with $\mathcal{V}_{\mathrm{sel}}$ and serves as the vocabulary basis for trajectory alignment.

\noindent\textit{(b) Vocabulary-guided trajectory alignment.}
LLaTSA performs cross-modal attention between $\mathbf{H}^{(0,k)}_{\mathrm{ts}}$ and the selected vocabulary embeddings to associate numerical trajectory patterns with pretrained textual representations. In this attention operation, the trajectory representation provides the queries, while the selected vocabulary embeddings provide the keys and values:
\begin{align}
\mathbf{Q}^{(k)}
&=
\mathbf{H}^{(0,k)}_{\mathrm{ts}} \cdot \mathbf{W}_{Q},
&
\mathbf{W}_{Q}
&\in \mathbb{R}^{d_{\mathrm{model}} \times d_k},
\\
\mathbf{K}
&=
\mathbf{E}_{\mathrm{sel}} \cdot \mathbf{W}_{K},
&
\mathbf{W}_{K}
&\in \mathbb{R}^{d_{\mathrm{model}} \times d_k},
\\
\mathbf{V}
&=
\mathbf{E}_{\mathrm{sel}} \cdot \mathbf{W}_{V},
&
\mathbf{W}_{V}
&\in \mathbb{R}^{d_{\mathrm{model}} \times d_{\mathrm{model}}},
\end{align}
where $\mathbf{Q}^{(k)}\in\mathbb{R}^{n_x\times N_p\times d_k}$, $\mathbf{K}\in\mathbb{R}^{|\mathcal{V}_{\mathrm{sel}}|\times d_k}$, and $\mathbf{V}\in\mathbb{R}^{|\mathcal{V}_{\mathrm{sel}}|\times d_{\mathrm{model}}}$. The attention weights characterize the correspondence between each temporal trajectory segment and the selected vocabulary entries. The aligned trajectory representation is obtained as
\begin{equation}
\mathbf{H}^{(0,k)}
=
\mathrm{softmax}\left(\frac{\mathbf{Q}^{(k)}\mathbf{K}^\top}{\sqrt{d_k}}\right)\mathbf{V}
\in \mathbb{R}^{n_x \times N_p \times d_{\mathrm{model}}}.
\label{eq:trajectory_alignment}
\end{equation}
The softmax term defines an attention matrix that associates each temporal trajectory segment with selected TSA-related vocabulary entries, and the weighted aggregation of the corresponding value vectors incorporates textual information into $\mathbf{H}^{(0,k)}$. The aligned representation therefore reduces the modality gap between numerical trajectory embeddings and the pretrained textual representation structure before subsequent trajectory prediction.

\subsection{Scenario-Aware Textual Prefix Encoding}
\label{sec:textual_prefix}

Vocabulary-guided alignment yields $\mathbf{H}^{(0,k)}$ for subsequent LLM-based trajectory prediction. In online TSA, however, a short post-fault trajectory segment may not contain sufficient information to distinguish cases with similar initial responses but different subsequent dynamic behavior, as discussed in \autoref{sec:short_observation_discrimination}. This subsection introduces a scenario-aware textual prefix that provides OC, disturbance, and variable-identity information in addition to $\mathbf{H}^{(0,k)}$. The prefix is generated from structured task, scenario, and variable records and prepended to the aligned trajectory representation, as shown in \autoref{fig:prompt_template}. The encoding procedure is described below.

\begin{figure*}[!t]
    \centering
    \begin{minipage}[t]{0.96\textwidth}
        \centering\footnotesize\textbf{(a) Prompt template}\par\smallskip
        \begin{promptframe}
        \footnotesize
        \textbf{Role:} Transient-dynamics trajectory predictor.\par\smallskip
        \textbf{Task:} Estimate the future post-fault trajectory of the specified state variable from short trajectory observations and scenario context.\par\smallskip
        \textbf{Scenario:} \textbf{System:} \textit{\{system\_name\}}. \textbf{Operating condition:} \textit{\{load\_level\}}, \textit{\{renewable\_penetration\}}. \textbf{Disturbance:} \textit{\{contingency\_type\}} \textit{\{fault\_type\}} at \textit{\{fault\_location\}}, cleared at \textit{\{clearing\_time\}}.\par\smallskip
        \textbf{Target state:} \textbf{Variable:} \textit{\{variable\_type\}}. \textbf{Location:} \textit{\{device\_or\_bus\_id\}}. \textbf{Unit:} \textit{\{unit\}}.
        \end{promptframe}
    \end{minipage}\par\medskip
    \begin{minipage}[t]{0.96\textwidth}
        \centering\footnotesize\textbf{(b) Instantiated example}\par\smallskip
        \begin{promptframe}
        \footnotesize
        \textbf{Role:} Transient-dynamics trajectory predictor.\par\smallskip
        \textbf{Task:} Estimate the future post-fault trajectory of the specified state variable from short trajectory observations and scenario context.\par\smallskip
        \textbf{Scenario:} \textbf{System:} IEEE 39-bus system. \textbf{Operating condition:} Load level 1.05 p.u., renewable penetration 40\%. \textbf{Disturbance:} N-2 three-phase short-circuit fault at buses 16 and 23, cleared at 0.10 s.\par\smallskip
        \textbf{Target state:} \textbf{Variable:} Rotor angle. \textbf{Location:} Generator G3. \textbf{Unit:} rad.
        \end{promptframe}
    \end{minipage}
    \caption{Textual-prefix template and instantiated example. The prompt combines task, scenario, and target-state information for operating-context conditioning.}
    \label{fig:prompt_template}
\end{figure*}

Accordingly, the auxiliary prefix for the $j$-th state variable of the $k$-th sample is defined as
\begin{equation}
\mathcal{C}_{\mathrm{aux}}^{(k,j)}
=
\big[
\mathcal{C}_{\mathrm{task}};
\mathcal{C}_{\mathrm{scen}}^{(k)};
\mathcal{C}_{\mathrm{var}}^{(j)}
\big],
\end{equation}
where the three components denote the task description, the OC-disturbance description of the $k$-th case, and the state-variable description, respectively.
The constructed prefix $\mathcal{C}_{\mathrm{aux}}^{(k,j)}$ is then tokenized using the pretrained LLM vocabulary and mapped to the backbone dimension:
\begin{equation}
\mathbf{E}_{\mathrm{aux}}^{(k,j)}
=
\mathrm{Tok}(\mathcal{C}_{\mathrm{aux}}^{(k,j)})\mathbf{E}_{\mathrm{vocab}}
\in \mathbb{R}^{L_{\mathrm{aux}} \times d_{\mathrm{model}}},
\end{equation}
where $\mathrm{Tok}(\mathcal{C}_{\mathrm{aux}}^{(k,j)}) \in \mathbb{R}^{L_{\mathrm{aux}} \times |\mathcal{V}|}$ is the token indicator matrix, and $\mathbf{E}_{\mathrm{vocab}} \in \mathbb{R}^{|\mathcal{V}| \times d_{\mathrm{model}}}$ is the pretrained vocabulary embedding table. The prefix representations for all state variables are stacked as $\mathbf{E}_{\mathrm{aux}}^{(k)}\in\mathbb{R}^{n_x\times L_{\mathrm{aux}}\times d_{\mathrm{model}}}$ and prepended to the aligned trajectory representation:
\begin{equation}
\mathbf{H}^{(0,k)}_{\mathrm{aug}}
=
\big[\,\mathbf{E}_{\mathrm{aux}}^{(k)} \,;\, \mathbf{H}^{(0,k)}\,\big]
\in \mathbb{R}^{n_x\times(L_{\mathrm{aux}} + N_p) \times d_{\mathrm{model}}}.
\end{equation}
The resulting $\mathbf{H}^{(0,k)}_{\mathrm{aug}}$ provides a scenario-conditioned trajectory representation for the subsequent LLM backbone.

\begin{figure}[t]
    \centering
    \includegraphics[width=\linewidth]{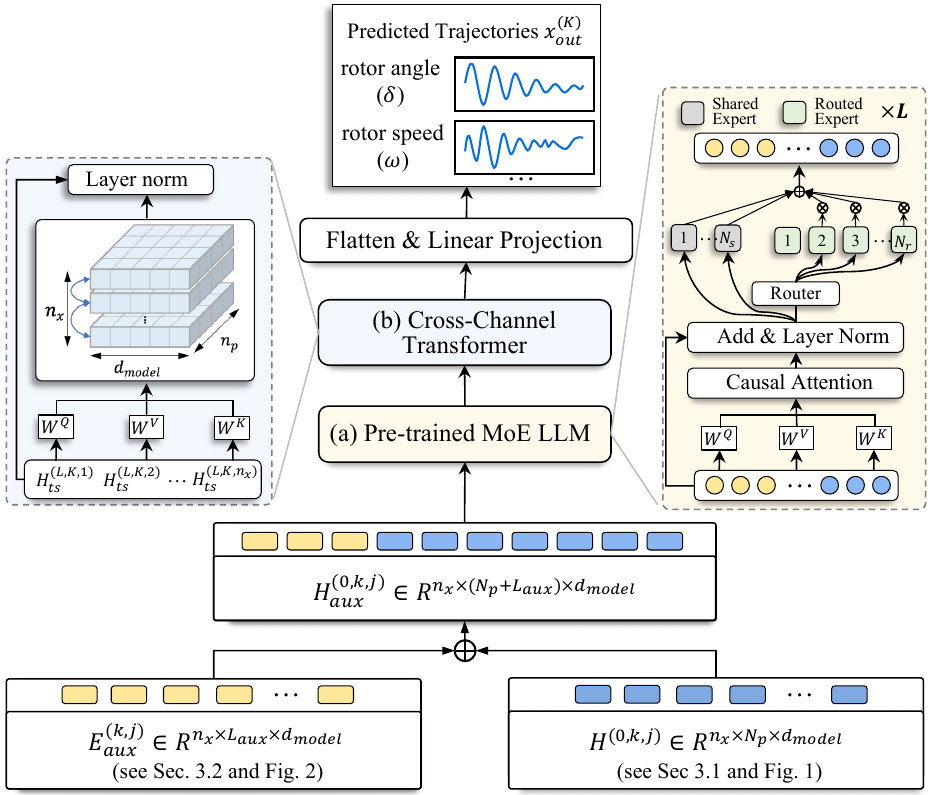}
    \caption{Backbone modeling and state variable coupling in the proposed framework. The augmented token representation is processed by (a) the pretrained MoE LLM backbone, where causal attention and sparse expert routing model autoregressive temporal dependencies, and (b) the variable coupling module, where information exchange among state variables captures coupled post-fault dynamics before trajectory prediction.}
    \label{fig:backbone_coupling}
\end{figure}

\subsection{MoE-Based Pretrained LLM Backbone for Dynamic Trajectory Modeling}\label{sec:moe_backbone}

This subsection specifies the pretrained MoE LLM backbone that maps $\mathbf{H}^{(0,k)}_{\mathrm{aug}}$ to temporal representations for trajectory prediction, as shown in \autoref{fig:backbone_coupling}(a). The backbone contains $L$ shared blocks, each consisting of causal self-attention over the scenario prefix and preceding trajectory segments, followed by an MoE feed-forward sublayer with sparse expert activation.

\subsubsection{Causal Self-Attention Mechanism}
For $l=1,\ldots,L$, let $\mathbf{H}_{\mathrm{aug}}^{(l-1,k)}$ denote the input to the $l$-th backbone block, with $\mathbf{H}_{\mathrm{aug}}^{(0,k)}$ as the initial input. The query, key, and value representations are
\begin{equation}
\begin{aligned}
\mathbf{Q}^{(l,k)} &= \mathbf{H}_{\mathrm{aug}}^{(l-1,k)}\cdot\mathbf{W}_Q^{(l)},\\
\mathbf{K}^{(l,k)} &= \mathbf{H}_{\mathrm{aug}}^{(l-1,k)}\cdot\mathbf{W}_K^{(l)},\\
\mathbf{V}^{(l,k)} &= \mathbf{H}_{\mathrm{aug}}^{(l-1,k)}\cdot\mathbf{W}_V^{(l)}.
\end{aligned}
\end{equation}
The causal attention output is
\begin{equation}
\mathrm{CSA}^{(l)}(\mathbf{H}_{\mathrm{aug}}^{(l-1,k)})
=
\mathrm{softmax}
\left(
\frac{\mathbf{Q}^{(l,k)}{\mathbf{K}^{(l,k)}}^\top}{\sqrt{d_k}}
+
\mathbf{M}
\right)
\mathbf{V}^{(l,k)} .
\end{equation}
Here, $\mathbf{M}\in\mathbb{R}^{L_{\mathrm{total}}\times L_{\mathrm{total}}}$ is the causal mask defined by
\begin{equation}
M_{a,b}
=
\begin{cases}
0, & b\le a,\\
-\infty, & b>a,
\end{cases}
\quad a,b=1,\ldots,L_{\mathrm{total}} .
\end{equation}
Since the scenario prefix is placed before the trajectory segments, the causal mask allows each trajectory segment to attend to the available scenario information and previous trajectory segments while blocking future trajectory segments. The residual connection and layer normalization then give
\begin{equation}
\widetilde{\mathbf{H}}_{\mathrm{aug}}^{(l,k)}
=
\mathrm{LayerNorm}\big(
\mathbf{H}_{\mathrm{aug}}^{(l-1,k)}
+
\mathrm{CSA}^{(l)}(\mathbf{H}_{\mathrm{aug}}^{(l-1,k)})
\big).
\label{eq:csa_update}
\end{equation}

\subsubsection{MoE Feed-Forward Network}
The MoE feed-forward sublayer updates $\widetilde{\mathbf{H}}_{\mathrm{aug}}^{(l,k)}$ through sparse expert routing. The following generic formulation accommodates checkpoint-specific architectures that use routed experts alone or combine routed experts with an always-active shared-expert set. The router computes the routed-expert probabilities for all state variables and sequence positions:
\begin{equation}
\boldsymbol{\Pi}^{(l,k)}
=
\mathrm{softmax}
\left(
\widetilde{\mathbf{H}}_{\mathrm{aug}}^{(l,k)}\mathbf{W}_g+\mathbf{b}_g
\right)
\in\mathbb{R}^{n_x\times L_{\mathrm{total}}\times N_r}.
\label{eq:moe_gating}
\end{equation}
At each position, only the top-$K$ routed experts are activated:
\begin{equation}
\mathcal{R}^{(l,k)}
=
\mathrm{Top}\text{-}K(\boldsymbol{\Pi}^{(l,k)}),
\end{equation}
where $\mathcal{R}^{(l,k)}$ stores the selected routed-expert indices. Let $\boldsymbol{\alpha}^{(l,k)}_e\in\mathbb{R}^{n_x\times L_{\mathrm{total}}}$ denote the normalized routing weight of routed expert $e$, with nonzero entries only where expert $e$ is selected. With $N_s$ shared experts and $N_r$ routed experts, the MoE update is
\begin{equation}
\begin{aligned}
\mathbf{H}_{\mathrm{aug}}^{(l,k)}
&=
\widetilde{\mathbf{H}}_{\mathrm{aug}}^{(l,k)}+
\sum_{s=1}^{N_s}
\mathrm{FFN}^{\mathrm{sh}}_{s}
\big(\widetilde{\mathbf{H}}_{\mathrm{aug}}^{(l,k)}\big)\\
&\quad+
\sum_{e=1}^{N_r}
\boldsymbol{\alpha}^{(l,k)}_e
\odot
\mathrm{FFN}^{\mathrm{rt}}_{e}
\big(\widetilde{\mathbf{H}}_{\mathrm{aug}}^{(l,k)}\big).
\end{aligned}
\label{eq:moe_update}
\end{equation}
Here, $\mathrm{FFN}^{\mathrm{sh}}_{s}(\cdot)$ and $\mathrm{FFN}^{\mathrm{rt}}_{e}(\cdot)$ denote the $s$-th shared expert and the $e$-th routed expert, respectively. The operator $\odot$ denotes element-wise weighting with broadcasting along the hidden dimension. For checkpoints without shared experts, $N_s=0$ and the shared-expert summation is omitted. Otherwise, shared experts remain active for all representations, whereas only the selected routed experts are activated at each position. After $L$ blocks, the backbone output $\mathbf{H}_{\mathrm{aug}}^{(L,k)}\in\mathbb{R}^{n_x\times L_{\mathrm{total}}\times d_{\mathrm{model}}}$ is passed to the state variable coupling module.

The MoE-based LLM backbone contributes to post-fault trajectory prediction in three ways. First, the decoder-only causal attention structure is consistent with online TSA, where future state evolution is estimated from OC, disturbance information, and the available post-fault trajectory history, without access to future trajectory segments. Second, the pretrained backbone provides a reusable temporal feature extractor for representing the evolution patterns of aligned rotor angle, rotor speed, and voltage trajectories. Third, the MoE feed-forward structure increases the capacity to represent diverse post-fault dynamic responses through shared and routed experts, while activating only a subset of routed experts at each position to control activated computation relative to dense scaling.

\subsection{State Variable Coupling for Dynamic Trajectory Prediction}
\label{sec:state_variable_coupling}

As discussed in \autoref{sec:coupled_state_evolution}, post-disturbance state variables are dynamically correlated and should be considered jointly for multivariable trajectory prediction. The preceding encoder and MoE backbone process each state variable with shared parameters, which keeps the model applicable to systems with different state dimensions. To incorporate variable dependence before prediction, LLaTSA reconstructs coupled state representations, as shown in \autoref{fig:backbone_coupling}(b).

The backbone output is $\mathbf{H}_{\mathrm{aug}}^{(L,k)} \in \mathbb{R}^{n_x\times L_{\mathrm{total}} \times d_{\mathrm{model}}}$. Because causal attention allows each trajectory patch to attend to the preceding textual prefix, the last $N_p$ patch representations contain both trajectory dynamics and operating-context information. They are used as the input to the coupling module:
\begin{equation}
\mathbf{C}^{(k,0)}
=
\mathbf{H}_{\mathrm{aug}}^{(L,k)}[:, -N_p:,:]
\in \mathbb{R}^{n_x\times N_p \times d_{\mathrm{model}}}.
\end{equation}
Let $\mathcal{P}$ denote a permutation operator that exchanges the state-variable and patch dimensions. The coupling input is rearranged as
\begin{equation}
\mathbf{S}^{(k,0)}
=
\mathcal{P}\big(\mathbf{C}^{(k,0)}\big)
\in \mathbb{R}^{N_p\times n_x\times d_{\mathrm{model}}}.
\end{equation}
In this representation, the state variable dimension becomes the attention dimension, and the same coupling operation can be evaluated for all trajectory patches in parallel. At the $m$-th layer of the state variable coupling module, the query, key, and value tensors are obtained as
\begin{align}
\mathbf{Q}^{(k,m)} &= \mathbf{S}^{(k,m-1)}\mathbf{W}_Q^{c,m}, \\
\mathbf{K}^{(k,m)} &= \mathbf{S}^{(k,m-1)}\mathbf{W}_K^{c,m}, \\
\mathbf{V}^{(k,m)} &= \mathbf{S}^{(k,m-1)}\mathbf{W}_V^{c,m},
\end{align}
where $\mathbf{Q}^{(k,m)}$, $\mathbf{K}^{(k,m)}$, and $\mathbf{V}^{(k,m)}$ have size $N_p\times n_x\times d_{\mathrm{model}}$. The trainable projection matrices satisfy
\begin{equation}
\mathbf{W}_Q^{c,m},\mathbf{W}_K^{c,m},\mathbf{W}_V^{c,m}
\in
\mathbb{R}^{d_{\mathrm{model}}\times d_{\mathrm{model}}}.
\end{equation}

The coupled representation is then computed by self-attention along the state variable dimension:
\begin{equation}
\widetilde{\mathbf{S}}^{(k,m)}
=
\mathrm{softmax}\left(
\frac{\mathbf{Q}^{(k,m)}(\mathbf{K}^{(k,m)})^{\top_x}}{\sqrt{d_{\mathrm{model}}}}
\right)
\mathbf{V}^{(k,m)},
\end{equation}
where $(\cdot)^{\top_x}$ denotes the transpose over the state variable and feature dimensions within each patch slice. The resulting attention tensor has size $N_p\times n_x\times n_x$, allowing each state variable to receive information from the other variables at the same patch position. Thus, the coupling module models local cross-variable dependence at each temporal patch, whereas the preceding backbone has already encoded temporal dependence within each state-variable sequence.

The representation is updated through residual integration and layer normalization:
\begin{equation}
\mathbf{S}^{(k,m)} =
\mathrm{LayerNorm}\big(
\mathbf{S}^{(k,m-1)} + \widetilde{\mathbf{S}}^{(k,m)}
\big).
\end{equation}
After $M$ coupling layers, the tensor is transformed back to the original variable and patch order:
\begin{equation}
\mathbf{C}^{(k,M)}
=
\mathcal{P}^{-1}\big(\mathbf{S}^{(k,M)}\big)
\in \mathbb{R}^{n_x\times N_p\times d_{\mathrm{model}}},
\end{equation}
which provides the multivariable trajectory representation for response prediction.

\subsection{Post-Fault Dynamic Response Prediction}
\label{sec:prediction_output}

After temporal modeling and state-variable coupling, $\mathbf{C}^{(k,M)}$ contains the post-disturbance representation of all state variables over the encoded trajectory patches. For future response prediction, the slice associated with the $c$-th state variable ($c=1,\ldots,n_x$) is first flattened along the patch and representation dimensions, and the first $d_{\mathrm{sel}}$ entries are retained:

\begin{equation}
\mathbf{z}_{\mathrm{sel}}^{(k,c)}
=
\mathcal{T}_{d_{\mathrm{sel}}}
\left(
\operatorname{flatten}\left(\mathbf{C}^{(k,M)}[c,:,:]\right)
\right)\in \mathbb{R}^{1\times d_{\mathrm{sel}}}.
\end{equation}
Here, $\operatorname{flatten}(\cdot)$ arranges the patch and representation dimensions into a row vector, and $\mathcal{T}_{d_{\mathrm{sel}}}(\cdot)$ denotes the truncation operation. The response of the $c$-th state variable over the prediction horizon is then obtained by
\begin{equation}
\widehat{\mathbf{x}}_{\mathrm{out}}^{(k,c)}
=
\left(
\mathbf{z}_{\mathrm{sel}}^{(k,c)}\mathbf{W}_{\mathrm{out}}
+ \mathbf{b}_{\mathrm{out}}
\right)^{\top}
\in \mathbb{R}^{L_{\mathrm{out}}},
\end{equation}
where $\mathbf{W}_{\mathrm{out}}\in\mathbb{R}^{d_{\mathrm{sel}}\times L_{\mathrm{out}}}$ and $\mathbf{b}_{\mathrm{out}}\in\mathbb{R}^{1\times L_{\mathrm{out}}}$ are shared by all state variables. Stacking the predicted responses gives the post-fault trajectory:
\begin{equation}
\begin{aligned}
\widehat{\mathbf{X}}_{\mathrm{out}}^{(k)}
&=
\big[
\widehat{\mathbf{x}}_{\mathrm{out}}^{(k,1)},\ldots,
\widehat{\mathbf{x}}_{\mathrm{out}}^{(k,n_x)}
\big]\in \mathbb{R}^{L_{\mathrm{out}}\times n_x}.
\end{aligned}
\end{equation}
Since the output parameters are not tied to $n_x$, the same output form can be used for systems with different numbers of state variables.

\section{Model Training and Online Application}

This section presents the training and online application of LLaTSA for trajectory-based TSA. The model performs rolling post-fault response prediction by appending each predicted segment to the available trajectory record and using the updated record to construct the next input window. To make training consistent with this iterative prediction process, LLaTSA develops a two-stage curriculum learning strategy consisting of teacher forcing and rollout-based training.

\subsection{Curriculum Training}

\noindent\textit{Stage I: Teacher Forcing.}
In the teacher-forcing stage, LLaTSA is conditioned on ground-truth trajectory samples. The loss is defined as
\begin{equation}
\mathcal{L}_{\mathrm{TF}}(\Theta)
=
\frac{1}{N_{\mathrm{tr}} n_x L_{\mathrm{out}}}
\sum_{k=1}^{N_{\mathrm{tr}}}
\left\|
\widehat{\mathbf{X}}_{\mathrm{out}}^{(k)}
-
\mathbf{X}_{\mathrm{out}}^{(k)}
\right\|_F^2 .
\end{equation}
This stage provides the initialized model for rollout-based training.

\noindent\textit{Stage II: Rollout-Based Training.}
Starting from the true input window, LLaTSA predicts one response segment at each rollout step, appends the prediction to the available trajectory record, and constructs the next input window from the latest $L_{\mathrm{in}}$ samples. For the $r$-th rollout step, let $\widehat{\mathbf{X}}_{\mathrm{out},r}^{(k)}\in\mathbb{R}^{L_{\mathrm{out}}\times n_x}$ and $\mathbf{X}_{\mathrm{out},r}^{(k)}\in\mathbb{R}^{L_{\mathrm{out}}\times n_x}$ denote the predicted and target response segments, respectively. The rollout loss is
\begin{equation}
\mathcal{L}_{\mathrm{RO}}(\Theta)
=
\frac{1}{N_{\mathrm{tr}} R n_x L_{\mathrm{out}}}
\sum_{k=1}^{N_{\mathrm{tr}}}
\sum_{r=1}^{R}
\left\|
\widehat{\mathbf{X}}_{\mathrm{out},r}^{(k)}
-
\mathbf{X}_{\mathrm{out},r}^{(k)}
\right\|_F^2 ,
\end{equation}
where $R$ is the number of rollout steps used during training. This stage directly penalizes the accumulated prediction error along the rolling trajectory and aligns training with online TSA application.

\subsection{Online Application for TSA}

After a disturbance is detected, the trained model $\Phi(\cdot;\Theta^{*})$ receives the available post-fault measurements and auxiliary textual information for online trajectory prediction. The measurements form the trajectory record, while the operating condition, disturbance attributes, and state-variable information are organized as $\mathcal{C}_{\mathrm{aux}}$ using the template in \autoref{sec:textual_prefix} and tokenized into $\mathbf{E}_{\mathrm{aux}}$. At the $r$-th rollout step, the latest $L_{\mathrm{in}}$ samples are used as the input window $\mathbf{X}_{\mathrm{in},r}$, and LLaTSA predicts
\begin{equation}
\widehat{\mathbf{X}}_{\mathrm{out},r}
=
\Phi\big(
\mathbf{X}_{\mathrm{in},r},
\mathbf{E}_{\mathrm{aux}};
\Theta^{*}
\big).
\end{equation}
The predicted segment is appended to the trajectory record, and the updated record provides the latest $L_{\mathrm{in}}$ samples for the next rollout step. Repeating this procedure yields the predicted post-fault dynamic trajectory for subsequent TSA evaluation, such as stability judgment, stability-margin estimation, or emergency-control screening.}

{\color{black}
\section{Case Studies}

\subsection{Test System and Data Preparation}

\subsubsection{Simulation Settings and Model Configuration}

The proposed framework is tested on the IEEE 39-bus, NETS--NYPS 68-bus, IEEE 118-bus, and Iceland 189-bus systems. Renewable-integrated cases are obtained by replacing selected SGs with DFIG-based wind units at penetration levels of 0\%, 10\%, 20\%, 30\%, and 40\%. To further examine converter-dominated dynamics, an IBR-modified IEEE 39-bus system is constructed with VSG-controlled photovoltaic generation at 20\% and 40\% penetration levels.
OCs are generated by varying the base load within $\pm 20\%$ and solving the corresponding optimal power flow in MATPOWER. Post-fault trajectories are obtained from time-domain simulations of three-phase-to-ground faults. For the DFIG-integrated systems, the contingency set covers $N$-$1$, $N$-$2$, and $N$-$3$ events with uniformly sampled fault locations. Each fault is applied at $t=1$ s, and its clearing time is randomly selected from 0--0.3 s after inception. The DFIG-integrated systems are simulated in PSAT for 4 s with a 0.02-s time step. For the IBR-modified system, a three-phase fault is applied at $t=0.55$ s and cleared at $t=0.56$ s. The Simulink simulation horizon is 1 s with a $10^{-5}$-s time step.

Unless otherwise specified, the conventional-system cases use $L_{\mathrm{in}}=65$ and $L_{\mathrm{out}}=1$. For the IBR-modified system, the EMT trajectories are temporally resampled from $10^{-5}$ s to $10^{-3}$ s and encoded with $L_{\mathrm{in}}=400$ and $L_{\mathrm{out}}=1$, providing a 0.40-s observation window after fault clearing while retaining tractable patch-token sequences. Temporal patching uses patch length $\tau=5$ and stride $s=3$ for all systems. The MoE LLM backbone is OLMoE-1B-7B with $d_{\mathrm{model}}=2048$, 16 decoder blocks, 64 experts, and 8 activated experts per token. Qwen1.5-MoE-A2.7B is also tested as an alternative backbone. The state-variable coupling module uses $M=3$ layers. LoRA is applied to selected decoder projections with rank $r=16$ and scaling coefficient $\alpha=32$. The model is trained using Adam with an initial learning rate of $1\times10^{-4}$ and batch size 32. Early stopping is based on validation loss, and training uses HuggingFace Accelerate, DeepSpeed ZeRO-2, and bf16 mixed precision. Each method is independently trained and evaluated using 20 distinct random seeds. All tabulated results are reported as the mean $\pm$ standard deviation across these 20 runs.

\subsubsection{Dataset Organization}

The trajectory dataset is organized according to three scenario dimensions: system configuration $\Xi_S$, renewable penetration level $\Xi_P$, and contingency order $\Xi_Q$. The complete scenario space is denoted by
\begin{equation}
\Xi
=
\Xi_S\times\Xi_P\times\Xi_Q,
\end{equation}
where each scenario instance $\xi=(s,p,q)\in\Xi$ specifies a power system configuration, a renewable penetration level, and a contingency condition. To distinguish mathematical system identifiers from benchmark names, the four conventional configurations are denoted by $S_{39}$, $S_{68}$, $S_{118}$, and $S_{189}$, corresponding to the IEEE 39-bus, NETS--NYPS 68-bus, IEEE 118-bus, and Iceland 189-bus systems, respectively. The IBR-modified IEEE 39-bus configuration is denoted by $S_{\mathrm{IBR}}$. Accordingly,
\begin{equation}
\Xi_S
=
\{S_{39},S_{68},S_{118},S_{189},S_{\mathrm{IBR}}\}.
\end{equation}
The renewable penetration and contingency dimensions are defined as
\begin{equation}
\Xi_P=\{0\%,10\%,20\%,30\%,40\%\},
\end{equation}
\begin{equation}
\Xi_Q=\{N\text{-}1,N\text{-}2,N\text{-}3\}.
\end{equation}
Accordingly, the full trajectory dataset can be written as
\begin{equation}
\mathcal{D}
=
\{D(\xi)\mid \xi\in\Xi\},
\end{equation}
where $D(\xi)$ denotes the set of trajectories generated under scenario $\xi$.

The training set deliberately excludes the two target-system configurations, one renewable penetration level, and one contingency order:
\begin{equation}
\begin{aligned}
\Xi_{\mathrm{train}}
=
&(\Xi_S\setminus\{S_{118},S_{\mathrm{IBR}}\})
\times(\Xi_P\setminus\{40\%\})\\
&\times
(\Xi_Q\setminus\{N\text{-}3\}).
\end{aligned}
\end{equation}
and
\begin{equation}
\mathcal{D}_{\mathrm{train}}
=
\{D(\xi)\mid \xi\in\Xi_{\mathrm{train}}\}.
\end{equation}
Thus, the IEEE 118-bus and IBR-modified IEEE 39-bus systems, 40\% renewable penetration, and $N$-$3$ contingencies are reserved for out-of-distribution evaluation. The validation set uses the same scenario range as $\mathcal{D}_{\mathrm{train}}$ but has disjoint OCs and fault locations. Test sets are defined in each case study according to the targeted generalization setting.

Prediction accuracy is evaluated using MAE and NRMSE. Stability discrimination is evaluated using unstable-class recall and balanced accuracy. For a test set with $N_{\mathrm{s}}$ samples and $n_x$ evaluated state channels,
\begin{align}
&\operatorname{MAE} =
\tfrac{1}{N_{\mathrm{s}} n_x L_{\mathrm{out}}}
\sum_{k=1}^{N_{\mathrm{s}}}
\sum_{c=1}^{n_x}
\left\|
\widehat{\mathbf{x}}_{\mathrm{out}}^{(k,c)}
-
\mathbf{x}_{\mathrm{out}}^{(k,c)}
\right\|_1, \\
&\operatorname{NRMSE} =
\tfrac{1}{n_x}
\sum_{c=1}^{n_x}
\tfrac{
\sqrt{
\tfrac{1}{N_{\mathrm{s}}L_{\mathrm{out}}}
\sum_{k=1}^{N_{\mathrm{s}}}
\left\|
\widehat{\mathbf{x}}_{\mathrm{out}}^{(k,c)}
-
\mathbf{x}_{\mathrm{out}}^{(k,c)}
\right\|_2^2
}
}
{
x_{\max}^{(c)}-x_{\min}^{(c)}
}.
\\
&\operatorname{Recall}_{\mathrm{u}} =
\frac{\mathrm{TP}_{\mathrm{u}}}
{\mathrm{TP}_{\mathrm{u}}+\mathrm{FN}_{\mathrm{u}}}, \\
&\operatorname{BAcc} =
\frac{1}{2}
\left(
\frac{\mathrm{TP}_{\mathrm{u}}}{\mathrm{TP}_{\mathrm{u}}+\mathrm{FN}_{\mathrm{u}}}
+
\frac{\mathrm{TN}_{\mathrm{s}}}{\mathrm{TN}_{\mathrm{s}}+\mathrm{FP}_{\mathrm{s}}}
\right).
\end{align}
Here, $x_{\max}^{(c)}$ and $x_{\min}^{(c)}$ are computed from the corresponding test trajectories of the $c$-th state channel and are used only to normalize the reported error. For stability discrimination, the ground-truth and predicted rotor-angle trajectories are respectively used to calculate $\mathrm{TSI}$ and $\widehat{\mathrm{TSI}}$. The resulting labels are defined as $y=\mathbb{I}(\mathrm{TSI}<0)$ and $\hat{y}=\mathbb{I}(\widehat{\mathrm{TSI}}<0)$, where the unstable class is treated as positive. $\mathrm{TP}_{\mathrm{u}}$ and $\mathrm{FN}_{\mathrm{u}}$ denote correctly detected and missed unstable cases, whereas $\mathrm{TN}_{\mathrm{s}}$ and $\mathrm{FP}_{\mathrm{s}}$ denote correctly and incorrectly classified stable cases. Thus, $\mathrm{Recall}_{\mathrm{u}}$ quantifies the detection rate of unstable cases, and $\mathrm{BAcc}$ gives equal weight to stable- and unstable-class recall under class imbalance.

\subsection{General-Purpose Dynamic Prediction Performance}
\label{sec:general_prediction}

This case evaluates general-purpose rotor-angle prediction across conventional benchmark systems under both in-range and out-of-distribution (OOD) scenario conditions. The overall test set is defined as
\begin{equation}
\mathcal{D}_{\mathrm{test}}
=
\{D(\xi)\mid \xi\in
(\Xi_S\setminus\{S_{118},S_{\mathrm{IBR}}\})
\times\Xi_P\times\Xi_Q
\},
\end{equation}
The IEEE 118-bus and IBR-modified systems are reserved for the cross-system and cross-dynamics study in \autoref{sec:cross_system_transfer}. Test samples within the training scenario range use OCs and fault locations disjoint from those used for training. The held-out $N$-$3$ contingency order and 40\% renewable penetration level are used to assess OOD scenario generalization. Rotor angle is selected as the representative dynamic state for the reported metrics. The IBR case additionally includes the VSG-equivalent rotor-speed response. Chronos~\cite{ansari2024chronos} and Timer~\cite{liu2024timer} are general-purpose time-series foundation models, whereas Uni-TSA~\cite{shen2025universal} is a general-purpose TSA baseline. These methods and LLaTSA are evaluated under identical data splits.

\begin{table}
\centering
\caption{Rotor-angle dynamic trajectory prediction performance on unseen operating conditions and fault locations within the training scenario range.}
\label{tab:general_dynamic_prediction_angle}
\renewcommand{\arraystretch}{1.05}
\setlength{\tabcolsep}{3pt}
\begin{tabularx}{\linewidth}{>{\centering\arraybackslash}m{0.19\linewidth}l>{\centering\arraybackslash}X>{\centering\arraybackslash}X>{\centering\arraybackslash}X>{\centering\arraybackslash}X}
\toprule
\textbf{System} & \textbf{Method} & \textbf{MAE} & \textbf{NRMSE} & \textbf{Rec. (\%)} & \textbf{BAcc. (\%)} \\
\midrule
\multirow{4}{*}{\makecell{IEEE\\39-bus}}
& Chronos & \meanstd{1.74}{0.18} & \meanstd{1.48}{0.12} & \meanstd{72.4}{3.6} & \meanstd{75.1}{3.2} \\
& Timer   & \meanstd{1.94}{0.21} & \meanstd{1.63}{0.15} & \meanstd{68.9}{3.4} & \meanstd{72.6}{3.1} \\
& Uni-TSA & \meanstd{0.58}{0.06} & \meanstd{0.58}{0.05} & \meanstd{86.7}{2.1} & \meanstd{88.1}{2.0} \\
\rowcolor{black!10}
& LLaTSA  & \meanstd{\textbf{0.14}}{0.02} & \meanstd{\textbf{0.25}}{0.03} & \meanstd{\textbf{97.8}}{1.0} & \meanstd{\textbf{98.1}}{0.9} \\
\midrule
\multirow{4}{*}{\makecell{NETS--NYPS\\68-bus}}
& Chronos & \meanstd{2.52}{0.27} & \meanstd{1.89}{0.16} & \meanstd{63.1}{4.2} & \meanstd{68.4}{3.8} \\
& Timer   & \meanstd{2.28}{0.24} & \meanstd{1.71}{0.14} & \meanstd{65.5}{4.0} & \meanstd{70.2}{3.7} \\
& Uni-TSA & \meanstd{0.94}{0.09} & \meanstd{0.77}{0.07} & \meanstd{82.0}{2.7} & \meanstd{84.9}{2.5} \\
\rowcolor{black!10}
& LLaTSA  & \meanstd{\textbf{0.20}}{0.03} & \meanstd{\textbf{0.34}}{0.03} & \meanstd{\textbf{96.5}}{1.2} & \meanstd{\textbf{97.2}}{1.1} \\
\midrule
\multirow{4}{*}{\makecell{Iceland\\189-bus}}
& Chronos & \meanstd{3.42}{0.34} & \meanstd{2.31}{0.21} & \meanstd{55.8}{4.8} & \meanstd{61.7}{4.3} \\
& Timer   & \meanstd{3.98}{0.41} & \meanstd{2.55}{0.24} & \meanstd{52.6}{5.0} & \meanstd{59.8}{4.5} \\
& Uni-TSA & \meanstd{1.49}{0.15} & \meanstd{1.01}{0.09} & \meanstd{78.4}{3.1} & \meanstd{81.6}{2.9} \\
\rowcolor{black!10}
& LLaTSA  & \meanstd{\textbf{0.32}}{0.04} & \meanstd{\textbf{0.45}}{0.04} & \meanstd{\textbf{94.2}}{1.5} & \meanstd{\textbf{95.1}}{1.4} \\
\midrule
\multirow{4}{*}{Average}
& Chronos & \meanstd{2.14}{0.22} & \meanstd{1.69}{0.16} & \meanstd{63.8}{4.2} & \meanstd{68.4}{3.8} \\
& Timer   & \meanstd{2.25}{0.23} & \meanstd{1.76}{0.18} & \meanstd{62.3}{4.1} & \meanstd{67.5}{3.8} \\
& Uni-TSA & \meanstd{0.83}{0.08} & \meanstd{0.71}{0.07} & \meanstd{82.4}{2.6} & \meanstd{84.9}{2.5} \\
\rowcolor{black!10}
& LLaTSA  & \meanstd{\textbf{0.20}}{0.03} & \meanstd{\textbf{0.31}}{0.03} & \meanstd{\textbf{96.2}}{1.2} & \meanstd{\textbf{96.8}}{1.1} \\
\bottomrule
\end{tabularx}
\end{table}

Table~\ref{tab:general_dynamic_prediction_angle} reports performance on unseen operating conditions and fault locations within the training scenario range. Chronos~\cite{ansari2024chronos} and Timer~\cite{liu2024timer} have substantially larger trajectory errors than the two TSA-oriented methods on every evaluated system. LLaTSA is the best-performing method across all four metrics. Relative to Uni-TSA, its average MAE decreases from 0.83 to 0.20 and its average NRMSE decreases from 0.71 to 0.31, while unstable-class recall and balanced accuracy increase from 82.4\% to 96.2\% and from 84.9\% to 96.8\%, respectively. The improvement in both trajectory error and trajectory-derived stability metrics shows that the gain is not confined to pointwise prediction accuracy. The ablation results in Table~\ref{tab:ablation} further quantify the contributions of scenario conditioning, vocabulary-guided alignment, and state-variable coupling to this performance.

\begin{table}
\centering
\caption{Rotor-angle prediction performance under unseen contingency order and renewable penetration level.}
\label{tab:unseen_contingency_res}
\renewcommand{\arraystretch}{1.08}
\setlength{\tabcolsep}{2pt}
\scriptsize
\begin{tabularx}{\linewidth}{>{\centering\arraybackslash}m{0.25\linewidth}l>{\centering\arraybackslash}X>{\centering\arraybackslash}X>{\centering\arraybackslash}X>{\centering\arraybackslash}X}
\toprule
\textbf{Test setting} & \textbf{Method} & \textbf{MAE} & \textbf{NRMSE} & \textbf{Rec. (\%)} & \textbf{BAcc. (\%)} \\
\midrule
\multirow{4}{*}{\makecell{Unseen $N$-$3$\\contingency}}
& Chronos & \meanstd{8.91}{0.89} & \meanstd{2.15}{0.21} & \meanstd{58.6}{4.7} & \meanstd{64.3}{4.2} \\
& Timer   & \meanstd{7.85}{0.78} & \meanstd{1.89}{0.19} & \meanstd{61.2}{4.5} & \meanstd{66.7}{4.0} \\
& Uni-TSA & \meanstd{3.92}{0.39} & \meanstd{1.15}{0.11} & \meanstd{79.8}{3.0} & \meanstd{82.7}{2.8} \\
\rowcolor{black!10}
& LLaTSA  & \meanstd{\textbf{0.94}}{0.09} & \meanstd{\textbf{0.51}}{0.05} & \meanstd{\textbf{95.6}}{1.3} & \meanstd{\textbf{96.4}}{1.2} \\
\midrule
\multirow{4}{*}{\makecell{Unseen\\40\% RES}}
& Chronos & \meanstd{6.81}{0.68} & \meanstd{1.69}{0.17} & \meanstd{62.4}{4.2} & \meanstd{67.1}{3.8} \\
& Timer   & \meanstd{7.15}{0.72} & \meanstd{1.77}{0.18} & \meanstd{59.7}{4.4} & \meanstd{65.5}{4.0} \\
& Uni-TSA & \meanstd{3.20}{0.32} & \meanstd{0.93}{0.09} & \meanstd{81.5}{2.8} & \meanstd{84.0}{2.6} \\
\rowcolor{black!10}
& LLaTSA  & \meanstd{\textbf{0.81}}{0.08} & \meanstd{\textbf{0.44}}{0.04} & \meanstd{\textbf{94.1}}{1.5} & \meanstd{\textbf{95.0}}{1.4} \\
\bottomrule
\end{tabularx}
\end{table}

Table~\ref{tab:unseen_contingency_res} evaluates scenario generalization at the held-out $N$-$3$ contingency order and 40\% renewable penetration level. LLaTSA remains the best-performing method across all four metrics in both settings. Its unstable-class recall exceeds 94\% and its balanced accuracy exceeds 95\%, despite the shift in contingency severity or renewable penetration. These results establish improved extrapolation to the two held-out scenario factors within the evaluated benchmark systems.}

\begin{figure*}
\centering
\includegraphics[width=\linewidth]{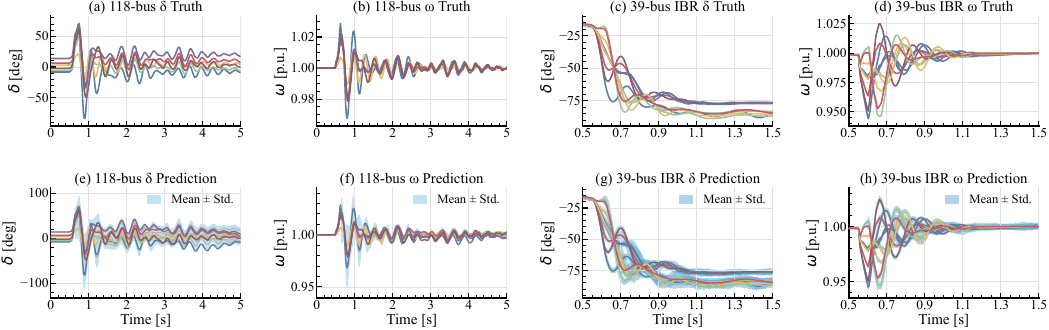}
\caption{Predicted and actual dynamic responses on the IBR-modified IEEE 39-bus system and the held-out IEEE 118-bus system.}
\label{fig:cross_system_prediction}
\end{figure*}

\begin{figure}
\centering
\includegraphics[width=0.8\linewidth]{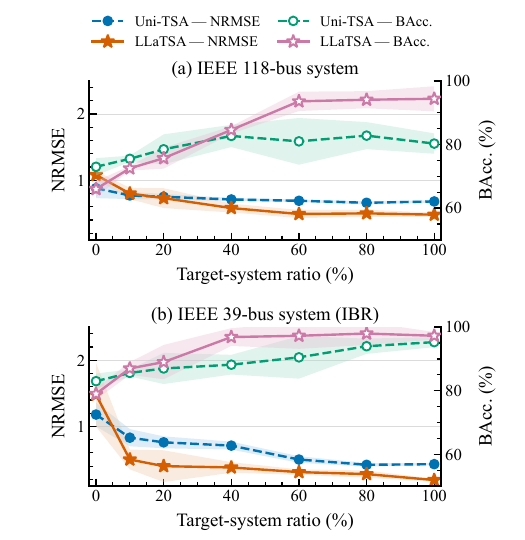}
\caption{Performance variation with increasing target-system data ratio during further fine-tuning for (a) the IBR-modified IEEE 39-bus system and (b) the held-out IEEE 118-bus system.}
\label{fig:target_system_data_scaling}
\end{figure}

\subsection{Cross-System and Cross-Dynamics Prediction}
\label{sec:cross_system_transfer}
The cross-system and cross-dynamics transfer capability of LLaTSA is evaluated on two target systems excluded from both training and validation. The IEEE 118-bus system represents an unseen network topology with a different system dimension and state-channel set. The IBR-modified IEEE 39-bus system retains the network topology but replaces part of the generation with VSG-controlled photovoltaic units. Its predicted states therefore include the VSG-equivalent rotor-angle and rotor-speed responses, and the case introduces converter-dominated dynamics in addition to the conventional generator responses. The corresponding test sets are
\begin{equation}
\mathcal{D}_{118}
=
\{D(\xi)\mid \xi\in\{S_{118}\}\times\Xi_P\times\Xi_Q\}.
\end{equation}
and
\begin{equation}
\mathcal{D}_{\mathrm{IBR}}
=
\{D(\xi)\mid \xi\in\{S_{\mathrm{IBR}}\}\times\{20\%,40\%\}\times\Xi_Q\}.
\end{equation}
The two settings distinguish generalization to an unseen network structure from generalization to an unseen dynamic regime and temporal resolution. They further assess whether the common backbone can be adapted with limited target-system data, rather than retrained separately for each system configuration.

To characterize adaptation efficiency, the fraction of target-system samples used for further fine-tuning is progressively increased. As shown in Fig.~\ref{fig:target_system_data_scaling}, both methods benefit from additional target-system data, but LLaTSA exhibits a faster reduction in NRMSE and a steeper, sustained increase in balanced accuracy. This progression is consistent with its scenario-conditioned, aligned representation providing a more transferable initialization. The scaling curves therefore establish the sample-efficiency advantage of LLaTSA before the two transfer mismatches are examined separately.

\begin{table}
\centering
\caption{Rotor-angle prediction on unseen target systems under limited and full target-data adaptation.}
\label{tab:cross_system_118}
\renewcommand{\arraystretch}{1.08}
\setlength{\tabcolsep}{2pt}
\scriptsize
\begin{tabularx}{\linewidth}{>{\centering\arraybackslash}p{0.12\linewidth}>{\raggedright\arraybackslash}p{0.26\linewidth}>{\centering\arraybackslash}X>{\centering\arraybackslash}X>{\centering\arraybackslash}X>{\centering\arraybackslash}X}
\toprule
\textbf{Data (\%)} & \textbf{Method} & \textbf{MAE} & \textbf{NRMSE} & \textbf{Rec. (\%)} & \textbf{BAcc. (\%)} \\
\midrule
\rowcolor{black!8}
\multicolumn{6}{c}{\textbf{IEEE 118-bus system}} \\
\midrule
\multirow{2}{*}{20} & Uni-TSA & \meanstd{1.42}{0.15} & \meanstd{0.80}{0.08} & \meanstd{76.0}{3.1} & \meanstd{79.4}{2.8} \\
& LLaTSA  & \meanstd{\textbf{0.58}}{0.06} & \meanstd{\textbf{0.72}}{0.07} & \meanstd{72.8}{2.4} & \meanstd{75.1}{2.1} \\
\cmidrule(lr){1-6}
\multirow{3}{*}{100} & Uni-TSA & \meanstd{1.20}{0.12} & \meanstd{0.68}{0.07} & \meanstd{79.0}{2.8} & \meanstd{81.8}{2.5} \\
& LLaTSA  & \meanstd{\textbf{0.41}}{0.04} & \meanstd{\textbf{0.51}}{0.05} & \meanstd{\textbf{93.7}}{1.6} & \meanstd{\textbf{94.8}}{1.4} \\
& LSTM (target-specific) & \meanstd{0.85}{0.06} & \meanstd{0.79}{0.07} & \meanstd{64.2}{2.5} & \meanstd{70.1}{2.2} \\
\midrule
\rowcolor{black!8}
\multicolumn{6}{c}{\textbf{IBR-modified IEEE 39-bus system}} \\
\midrule
\multirow{2}{*}{20} & Uni-TSA & \meanstd{1.30}{0.13} & \meanstd{0.75}{0.08} & \meanstd{86.1}{2.8} & \meanstd{88.7}{2.5} \\
& LLaTSA  & \meanstd{\textbf{0.37}}{0.04} & \meanstd{\textbf{0.40}}{0.04} & \meanstd{88.9}{1.8} & \meanstd{90.1}{1.6} \\
\cmidrule(lr){1-6}
\multirow{3}{*}{100} & Uni-TSA & \meanstd{0.76}{0.08} & \meanstd{0.45}{0.05} & \meanstd{91.3}{2.5} & \meanstd{94.1}{2.2} \\
& LLaTSA  & \meanstd{\textbf{0.19}}{0.02} & \meanstd{\textbf{0.20}}{0.02} & \meanstd{\textbf{95.2}}{1.2} & \meanstd{\textbf{96.0}}{1.1} \\
& LSTM (target-specific) & \meanstd{0.78}{0.06} & \meanstd{0.62}{0.06} & \meanstd{85.3}{1.9} & \meanstd{84.5}{1.7} \\
\bottomrule
\end{tabularx}
\end{table}

Table~\ref{tab:cross_system_118} contrasts 20\% and full-data adaptation with a target-specific LSTM trained from scratch on all available target data. With only 20\% target-system data, LLaTSA surpasses the full-data LSTM on every reported metric for both target systems. On the held-out IEEE 118-bus system, for example, LLaTSA obtains an MAE of 0.58, an NRMSE of 0.72, an unstable-class recall of 72.8\%, and a balanced accuracy of 75.1\%, versus 0.85, 0.79, 64.2\%, and 70.1\%, respectively, for the full-data LSTM. This comparison demonstrates the value of mixed-system pretraining for data-efficient target-system adaptation. At the same 20\% ratio, LLaTSA has lower MAE and NRMSE than Uni-TSA but lower balanced accuracy (75.1\% versus 79.4\%). As shown in Fig.~\ref{fig:target_system_data_scaling}, this initial decision-metric gap reverses by 40\% target data and then widens: LLaTSA reaches a balanced accuracy of 94.8\% at full adaptation, compared with 81.8\% for Uni-TSA. The examples in Fig.~\ref{fig:cross_system_prediction} are consistent with the aggregate metrics, showing accurate reconstruction of both the initial rotor-angle excursion and the subsequent recovery. Thus, after full target-system adaptation, the advantage on the unseen network extends from trajectory accuracy to stability discrimination.

The IBR-modified 39-bus system poses a complementary dynamic mismatch. VSG-controlled photovoltaic generation introduces fast control modes absent from the DFIG-integrated training systems. A three-phase fault is applied at 0.55 s and cleared at 0.56 s within a 1-s electromagnetic-transient simulation that uses a $10^{-5}$-s step. The resulting trajectories are resampled at $10^{-3}$ s for model input. A longer $L_{\mathrm{in}}=400$ window preserves 0.40 s of the available 0.44-s post-clearing response, whereas the conventional-system cases use $L_{\mathrm{in}}=65$. Here, LLaTSA with only 20\% target data also surpasses the full-data target-specific LSTM, reducing MAE and NRMSE from 0.78 and 0.62 to 0.37 and 0.40, respectively, while increasing unstable-class recall and balanced accuracy from 85.3\% and 84.5\% to 88.9\% and 90.1\%. With 100\% target data, LLaTSA further outperforms Uni-TSA in MAE and NRMSE (0.19 versus 0.76 and 0.20 versus 0.45, respectively) and in unstable-class recall and balanced accuracy (95.2\% versus 91.3\% and 96.0\% versus 94.1\%, respectively). Consistent with the scaling results, Fig.~\ref{fig:cross_system_prediction} shows that LLaTSA captures the dominant excursion, damping, and recovery in the presence of fast control-induced fluctuations. The results across the two targets support adaptation of LLaTSA under both structural and dynamic mismatch.

\subsection{Computational Cost for Online TSA}
\label{sec:computational_cost}

This experiment compares the empirical accuracy--latency scaling of dense and sparse backbones under an identical prediction protocol. Uni-TSA uses GPT-2 backbones from 0.12B to 6B parameters. LLaTSA uses a common OLMoE-1B-7B checkpoint truncated to the first 1, 2, 3, 6, 14, or 16 decoder blocks, corresponding to approximately 0.63B, 1.06B, 1.48B, 2.76B, 6.15B, and 7.00B total parameters. All configurations use the same hardware, numerical precision, batch size, input window, and rollout horizon. NRMSE and end-to-end time per operating condition are reported.

\begin{figure}[!t]
\centering
\includegraphics[width=0.8\linewidth]{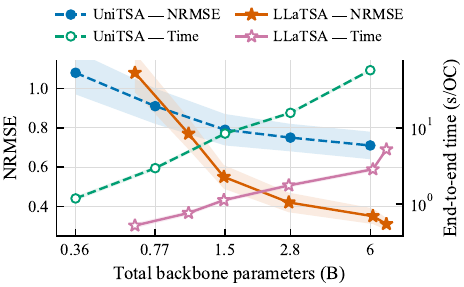}
\caption{Prediction accuracy and end-to-end inference time under backbone scaling. Solid curves show NRMSE and dashed curves show the time required to predict one complete trajectory for an operating condition. Shaded regions denote one standard deviation over repeated runs.}
\label{fig:backbone_scaling_cost}
\end{figure}

Figure~\ref{fig:backbone_scaling_cost} shows that both methods benefit from increased capacity, whereas the dense Uni-TSA backbone exhibits a steeper latency increase. At the largest configurations, LLaTSA attains an NRMSE of 0.31, consistent with Table~\ref{tab:general_dynamic_prediction_angle}, compared with 0.71 for Uni-TSA, while retaining substantially lower end-to-end time. Because the horizontal axis reports total parameters whereas an MoE activates only a subset of routed experts per token, the figure should be interpreted as a measured latency comparison under the stated protocol rather than as a parameter-count-normalized comparison. Under this protocol, LLaTSA provides a better empirical accuracy--latency tradeoff.

\subsection{Ablation Study and Backbone Comparison}

\label{sec:ablation}

\begin{table*}[!t]
\centering
\caption{Ablation results for the proposed representation. OOD results are averaged over the unseen $N$-$3$ contingency and 40\% renewable-penetration settings. The IEEE 118-bus and IBR 39-bus results use 20\% target-system data for further fine-tuning.}
\label{tab:ablation}
\renewcommand{\arraystretch}{1.08}
\setlength{\tabcolsep}{3.2pt}
\ifdefined\apenablationtablefont
\setlength{\tabcolsep}{2.6pt}
\apenablationtablefont
\else
\scriptsize
\fi
\begin{tabularx}{\textwidth}{>{\raggedright\arraybackslash}p{0.19\textwidth}>{\centering\arraybackslash}X>{\centering\arraybackslash}X>{\centering\arraybackslash}X>{\centering\arraybackslash}X>{\centering\arraybackslash}X>{\centering\arraybackslash}X>{\centering\arraybackslash}X>{\centering\arraybackslash}X}
\toprule
\multirow{2}{*}{\textbf{Variant}} & \multicolumn{2}{c}{\textbf{In-range}} & \multicolumn{2}{c}{\textbf{OOD}} & \multicolumn{2}{c}{\textbf{IEEE 118-bus, 20\%}} & \multicolumn{2}{c}{\textbf{IBR 39-bus, 20\%}} \\
\cmidrule(lr){2-3}\cmidrule(lr){4-5}\cmidrule(lr){6-7}\cmidrule(lr){8-9}
& \makecell{\textbf{NR}\\\textbf{MSE}} & \textbf{BAcc. (\%)} & \makecell{\textbf{NR}\\\textbf{MSE}} & \textbf{BAcc. (\%)} & \makecell{\textbf{NR}\\\textbf{MSE}} & \textbf{BAcc. (\%)} & \makecell{\textbf{NR}\\\textbf{MSE}} & \textbf{BAcc. (\%)} \\
\midrule
Uni-TSA & \meanstd{0.71}{0.07} & \meanstd{84.9}{2.6} & \meanstd{1.04}{0.10} & \meanstd{83.4}{2.9} & \meanstd{0.80}{0.08} & \textbf{\meanstd{79.4}{2.8}} & \meanstd{0.75}{0.08} & \meanstd{88.7}{2.5} \\
LLaTSA w/o alignment & \meanstd{0.43}{0.04} & \meanstd{94.1}{1.5} & \meanstd{0.68}{0.07} & \meanstd{91.2}{1.9} & \meanstd{0.88}{0.09} & \meanstd{64.2}{2.3} & \meanstd{0.69}{0.07} & \meanstd{86.4}{2.0} \\
LLaTSA w/o prefix & \meanstd{0.39}{0.04} & \meanstd{94.9}{1.4} & \meanstd{0.63}{0.06} & \meanstd{92.1}{1.8} & \meanstd{0.80}{0.08} & \meanstd{68.1}{2.2} & \meanstd{0.63}{0.06} & \meanstd{87.3}{1.9} \\
LLaTSA w/o coupling & \meanstd{0.36}{0.03} & \meanstd{95.6}{1.2} & \meanstd{0.55}{0.06} & \meanstd{94.0}{1.5} & \meanstd{0.75}{0.08} & \meanstd{70.8}{2.1} & \meanstd{0.60}{0.06} & \meanstd{88.5}{1.8} \\
LLaTSA w/o prefix and alignment & \meanstd{0.55}{0.05} & \meanstd{90.0}{2.0} & \meanstd{0.88}{0.09} & \meanstd{86.7}{2.5} & \meanstd{1.02}{0.10} & \meanstd{58.5}{2.8} & \meanstd{0.78}{0.08} & \meanstd{84.3}{2.4} \\
\rowcolor{black!8}
Full LLaTSA & \textbf{\meanstd{0.31}{0.03}} & \textbf{\meanstd{96.8}{1.1}} & \textbf{\meanstd{0.48}{0.05}} & \textbf{\meanstd{95.7}{1.4}} & \textbf{\meanstd{0.72}{0.07}} & \meanstd{75.1}{2.1} & \textbf{\meanstd{0.40}{0.04}} & \textbf{\meanstd{90.1}{1.6}} \\
\bottomrule
\end{tabularx}
\end{table*}

Table~\ref{tab:ablation} evaluates the contribution of the three representation components. ``w/o alignment'' replaces vocabulary-guided alignment with direct linear projection, ``w/o prefix'' removes the operating-condition, disturbance, and variable-identity prefix, and ``w/o coupling'' bypasses the state-variable coupling module. The combined variant removes both input-side components while retaining the common backbone and coupling module. The same training protocol is used for all LLaTSA variants.

Removing any component reduces trajectory accuracy and stability-discrimination performance. Vocabulary-guided alignment has the largest individual effect on NRMSE, increasing it from 0.31 to 0.43 in the in-range setting and from 0.48 to 0.68 under OOD conditions when removed. Removing the textual prefix produces comparable degradation under OOD and limited-data transfer, while bypassing state-variable coupling yields smaller but consistent losses, particularly in the OOD and IBR settings. Removing both input-side components produces the largest deterioration on the target systems, increasing IEEE 118-bus NRMSE from 0.72 to 1.02 and reducing IBR balanced accuracy from 90.1\% to 84.3\%. Together, these controlled comparisons show that the three components make complementary contributions in the evaluated settings.

The backbone comparison further shows that LLaTSA benefits from a stronger pretrained MoE prior. Replacing OLMoE-1B-7B with Qwen1.5-MoE-A2.7B further improves performance across the evaluated settings, particularly under distribution shift and limited-data transfer. These gains show that the proposed scenario conditioning, vocabulary-guided alignment, and state-variable coupling remain effective across pretrained MoE backbones, while a stronger backbone provides additional capacity for modeling heterogeneous post-fault dynamics.

\subsection{Interpretability Analysis}
\label{sec:interpretability}

\begin{figure}
\centering
\includegraphics[width=\linewidth]{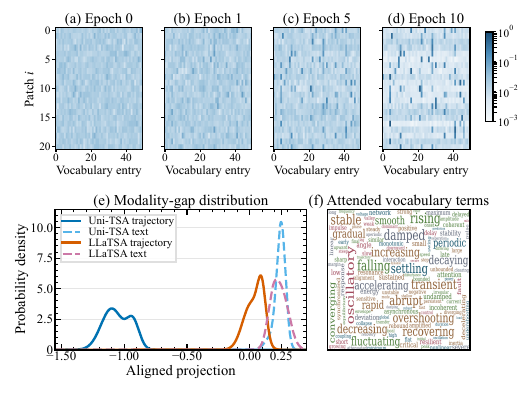}
\caption{Interpretability analysis of vocabulary-guided trajectory alignment. (a)--(d) Evolution of the patch-to-vocabulary attention matrix $\mathbf{A}^{(k,j)}$ during training. (e) Projected trajectory and text representation densities within the Uni-TSA and LLaTSA backbone spaces. The smaller within-backbone separation for LLaTSA is consistent with improved modality alignment. (f) Decoded vocabulary entries ranked by aggregate attention, where term size is proportional to the corresponding attention score.}
\label{fig:attention_modality}
\end{figure}

Fig.~\ref{fig:attention_modality} examines the learned behavior of vocabulary-guided trajectory alignment. For state channel $j$, the softmax term in \eqref{eq:trajectory_alignment} defines the patch-to-vocabulary attention matrix $\mathbf{A}^{(k,j)}$. In Figs.~\ref{fig:attention_modality}(a)--\ref{fig:attention_modality}(d), initially diffuse attention becomes concentrated at channel- and patch-dependent locations during training. The encoder therefore uses the candidate vocabulary nonuniformly when representing local post-fault response segments, rather than uniformly averaging all embeddings. This observation is consistent with selective encoding of the early response features relevant to subsequent trajectory evolution.

Figure~\ref{fig:attention_modality}(e) provides complementary representation-level evidence. After $L_2$ normalization and projection onto the trajectory--text center directions, the within-backbone separation between numerical trajectory and textual representations is smaller for LLaTSA than for Uni-TSA. Because the two methods use different native embedding spaces and vocabularies, only this within-backbone comparison is meaningful. The high-attention entries in Fig.~\ref{fig:attention_modality}(f) describe rising, falling, oscillatory, damped, settling, and recovering behavior, all of which are recognizable response motifs in post-fault rotor-angle and rotor-speed trajectories. The analysis does not assign a unique physical label to any token; instead, it is consistent with the alignment module placing numerical response patches in a representation space better matched to the pretrained backbone.

\section{Conclusion}

This paper introduces LLaTSA, a general-purpose framework that leverages a pretrained mixture-of-experts LLM for trajectory-based transient stability analysis across diverse operating conditions, contingencies, and power-system configurations. The framework retains channel-independent processing to accommodate variable state dimensions, while a scenario-aware textual prefix supplies operating-condition, disturbance, and state-variable information that cannot be reliably inferred from a short post-fault record alone. Numerical trajectory patches are then aligned with a compact TSA-related vocabulary before backbone prediction, reducing the representation mismatch between continuous dynamic responses and text-pretrained embeddings. Sparse expert routing and state-variable coupling subsequently model the temporal evolution and coordinated post-fault responses with controlled activated computation. Compared with Uni-TSA, LLaTSA reduces the average NRMSE from 0.71 to 0.31 and increases balanced accuracy from 84.9\% to 96.8\%. On both held-out target systems, LLaTSA adapted with only 20\% target-system data surpasses a target-specific LSTM trained from scratch on the complete target dataset, demonstrating sample-efficient adaptation under structural and dynamic mismatch. These results establish LLaTSA as a promising approach for online TSA in heterogeneous power systems.

\bibliographystyle{IEEEtran}
\bibliography{ref}

\end{document}